\documentclass[aps,twocolumn,prd]{revtex4}

\usepackage{latexsym}
\usepackage{epsfig}

\def\be{\begin{equation}}
\def\ee{\end{equation}}
\def\bea{\begin{eqnarray}}
\def\eea{\end{eqnarray}}
\def\ba{\begin{array}} 
\def\ea{\end{array}}
\def\bc{\begin{center}}
\def\ec{\end{center}}

\def\ghost#1{}

\def\simge{\mathrel{%
   \rlap{\raise 0.511ex \hbox{$>$}}{\lower 0.511ex \hbox{$\sim$}}}}
\def\simle{\mathrel{
   \rlap{\raise 0.511ex \hbox{$<$}}{\lower 0.511ex \hbox{$\sim$}}}}

\def\dis{\displaystyle}

\begin{document}

\title{Invisible $\Upsilon$ decays into Light Dark Matter
\vspace{5mm}\\
}

\author{P. Fayet\,$^1$
\vspace{5mm}
\\
$^1$ Laboratoire de Physique Th\'eorique de l'ENS
{\small \ (UMR 8549 CNRS)} \\
 24 rue Lhomond, 75231 Paris Cedex 05, France
\medskip \\
}


\begin{abstract}
{
\vspace{3mm}
Invisible $\psi$ and $\Upsilon$ decays into light neutralinos, within the MSSM or N(n)MSSM,
are smaller than for $\nu\bar\nu$ production, even if light spin-0 particles
are coupled to quarks and neutralinos.
In a more general way, light dark matter particles are normally forbidden, unless they can annihilate sufficiently through a new interaction stronger than weak interactions (at lower energies), as induced by a light spin-1 $\,U$ boson,
or heavy-fermion exchanges in the case of scalar dark matter.
We discuss the possible contributions of $\,U$-boson, heavy-fermion, or spin-0 exchanges to invisible $\psi$ and
$\Upsilon$ decays.
\vspace{.5mm}
 \\
\hspace*{2mm}
$U$-exchanges could lead, but not necessarily, to significant branching fractions 
for invisible decays
into light dark matter.\,\,We show how one can get the correct relic density together with sufficiently small invisible branching fractions, and the resulting constraints 
on the $\,U$ couplings to ordinary particles and dark matter, in particular $\,|\,c_\chi\,f_{bV}\,|<\,5\ 10^{-3}\,$
from $\Upsilon$ decays, for $2\,m_\chi$ smaller than a few GeV.
We also explain why there is no model-independent way to predict $\psi$ and $\Upsilon$ 
branching fractions into light dark matter, from dark matter annihilation cross sections
at freeze-out time.
\vspace{3.5mm}
\hfill \hbox{LPTENS/09-33}
\vspace{0mm}\\
}
\end{abstract}

\maketitle

\section{\vspace{1.5mm}
\boldmath New particles in invisible $\,\psi\,$ and $\,\Upsilon\,$ decays.}

\label{sec:intro}

\vspace{-1mm}

The nature of dark matter is one of the most challenging issues facing physics.
Observation of standard model (SM) particles coupling to invisible
final states, as searched for recently in $\Upsilon$ decays \cite{babarinv}, might provide information on new neutral particles such as photinos 
or neutralinos and very light gravitinos, and candidate dark matter constituents \cite{fayetpsi,fayetkaplan,prd06}.
In the standard model, invisible decays of the $\Upsilon(1S)$
proceed by $b\,\bar b$ annihilation into a $\,\nu\bar \nu$ pair,
with a branching fraction~\cite{fayetkaplan}
\be
\label{upsnu0}
B\,(\Upsilon(1S)\,\to\,\nu\,\bar\nu)\ \simeq \ 10^{-5}\ \ ,
\ee
well below the current experimental sensitivity \cite{babarinv}.
However, low-mass dark matter candidates could couple through {\it stronger-than-weak\,} interactions to SM particles, and possibly enhance the 
invisible branching fraction  of the $\Upsilon(1S)$ to the level of $\,\approx 10^{-5}$ to $\,10^{-2}$ \cite{fayetkaplan,prd06}, in contrast with weakly-interacting particles, as indicated  by 
(\ref{upsnu0}).
A new light boson $U$ associated with the gauging of an extra-$U(1)$ symmetry, as 
considered long ago in \cite{U}, may play a crucial role as a mediator of such a new interaction.

\vspace{2mm}

Upper limits on the invisible $\Upsilon$ branching fraction have been obtained long ago by CLEO  \cite{upsinv} and ARGUS \cite{argus}, already having in mind the search for new weakly-interacting particles such as photinos and very light gravitinos, as in invisible $\psi$ decays
\cite{fayetpsi}.
These limits ($7\ 10^{-3}$ for $\,\psi$, $\,5\ 10^{-2}$ then $\,2.3\ 10^{-2}$ for $\,\Upsilon$)
are obtained by looking for
\be
\left\{\ 
\ba{ccc}
\psi\,(2S) &\rightarrow& \ \pi^+\,\pi^-\ \psi\,(1S)_{\ \hookrightarrow\  \hbox{\footnotesize invisible}}\ \ ,
\vspace{2mm}\\
\Upsilon(3S),\ \hbox{or}\ \Upsilon(2S) &\rightarrow& \ \pi^+\,\pi^-\ \Upsilon(1S)_{\ \hookrightarrow\  \hbox{\footnotesize invisible}}\ \ ,
\ea \right. 
\ee
which provide signatures for the production and invisible decays of $\,\psi$ and $\,\Upsilon$.

\vspace{1.5mm}

The $\Upsilon$ bounds have been improved by Belle and CLEO \cite{upsinv2,cleo2}, and recently 
B{\small A}B{\small A}R \cite{babarinv}, down to
\be
\label{limups}
B\,(\,\Upsilon(1S) \ \to\ \hbox{invisible}\,)\ < \ 3\ 10^{-4}\ \ ,
\ee
at the 90\,\% c.l.\,.
We also have, from BES II \cite{bes},
\be
\label{limpsi}
B\,(\,\psi(1S) \ \to\ \hbox{invisible}\,)\ < \ 7.2\ 10^{-4}\ \ .
\ee
Although the present paper is in general formulated with the $\Upsilon$, 
the analysis applies to invisible $\psi\,$ decays as well.

\vspace{1.5mm}

What can we learn about the light neutral particles that could be produced\,?
We shall discuss possible invisible decays of the $\Upsilon$, 
\,and at first  into {\it light neutralinos} within the MSSM or 
N(n)MSSM (cf.~Sec.\,\ref{sec:ssm}). They are significantly
smaller than for $\Upsilon(1S)$ $\to\,\nu\bar\nu$, \,even in the presence of light spin-0 particles
coupling directly quarks to neutralinos.

\vspace{1.5mm}
We shall also discuss, in a more general way, invisible decays of $\,\psi\,$ and $\Upsilon\,$ 
into {\it \,light dark matter particles} 
(of mass $<m_\psi/2$ \,or $\, m_\Upsilon/2$) \cite{fayetkaplan,prd06},
which could be much {\it more strongly coupled to ordinary particles than through ordinary weak interactions}, possibly leading to significant invisible $\,\psi$ and $\,\Upsilon\,$ branching fractions.

\vspace{2mm}
Indeed, light dark matter (LDM) particles \cite{bf,fermion} are normally required to annihilate through a new interaction stronger than weak interactions (at least at lower energies), otherwise their relic density would be too large.
Is this compatible with the new experimental bounds  \cite{babarinv,bes} on  invisible $\psi$ and $\Upsilon$ decays\,? 
This is part of the more general question \cite{prd06,fermion,prd07}:
how can we have a new interaction stronger than weak interactions, responsible for sufficient annihilations 
of LDM particles in the early Universe,
and at the same time how could it remain unnoticed if it is stronger than weak interactions\,?
Indeed this stronger-than-weak feature cannot persist up to high energies, 
especially with (production, annihilation or interaction) cross sections increasing like $s$, 
without getting in conflict with experimental results. 

\vspace{1.75mm}

The apparent contradiction is solved for an interaction mediated by {\it a new light neutral boson} 
with small couplings to quarks and leptons, such as the light \hbox{spin-1} $U$ boson \cite{U} introduced and discussed long ago, associated with the gauging of an extra-$U(1)$ symmetry. 
Another possibility, for spin-0 dark matter particles, is obtained for interactions mediated by {\it heavy-fermion exchanges} \cite{bf}, which may also allow for sufficient annihilations of light dark matter in the early Universe.
We shall also discuss (in Sec.\,\ref{sec:ssm}) light spin-0 exchanges, which do not contribute to invisible
$\,\psi$ and $\Upsilon$ decays, both in the N(n)MSSM and in a more general way.

\vspace{1.75mm}
The choice is thus, for the production of light dark matter particles
in invisible $\,\psi$ and $\Upsilon$ decays,  between 
\be
\hbox{\it a new neutral current} \ \ \  \hbox{and\ \ \it \,new heavy fermions}\,,
\ee
or both
(reminding us of the early days of gauge theories, 
before the discovery of the weak neutral current coupled to the $Z$).
We shall discuss in Sec.~\ref{sec:scal} the possible production of scalar dark matter through heavy-fermion exchanges, according to

\vspace{-7mm}
\be
\Upsilon\ \ \stackrel{\ba{c}\hbox{\small $b_M$}\vspace{0mm}\\ \ea}{\longrightarrow}\ \ \varphi\,\bar\varphi\ \ .
\ee
We shall then concentrate, in Secs.~\ref{sec:U} and \ref{sec:comp}, on $U$-in\-duced reactions, discussing the implications of 
the experimental limit (\ref{limups}) on the couplings of 
the light $\,U$ boson 
 that may be responsible for $\,\Upsilon\,$ annihilations into light spin-$\frac{1}{2}$ 
\,($\chi$)\, or spin-0 \,($\varphi$)\, dark matter particles:
\be
\Upsilon\ \ \stackrel{\hbox{\footnotesize $U$}}{\rightarrow}\ \ \chi\,\chi\ \ \ \hbox{ (or $\,\varphi\,\bar\varphi\,$ or $\,\chi\,\bar\chi$)}\ \ ,
\ee
at $\,\sqrt s = m_\Upsilon$ 
\cite{prd06}. This $\,U$ allows for the correct relic density of light dark matter particles \cite{bf,fermion}, 
by inducing sufficient annihilations in the early Universe, 
\be
\chi\,\chi\ \ \ \hbox{(or $\,\varphi\,\bar\varphi\,$ or $\,\chi\,\bar\chi$)}
\ \ \stackrel{\hbox{\footnotesize $U$}}{\rightarrow}\ \ f\,\bar f\ \ ,
\ee
at lower values of the energy, $\,\sqrt s\,\simeq\,2\,m_\chi\,$ (or $2\,m_\varphi$).

\vspace{1.75mm}

Finally we shall discuss in Sec.\,\ref{sec:rel} whether it makes sense to attempt predicting, in a model-independent way,
 $\,\psi\,$ and $\,\Upsilon\,$ invisible branching fractions into dark matter, from dark-matter annihilation cross sections at freeze-out time; and consider briefly, in Sec.\,\ref{sec:npert}, possible non-perturbative effects associated with light $U$ exchanges.

\vspace{-.5mm}

\section{\vspace{1.2mm}
Invisible decays into neutrinos, 
\vspace{1.5mm}
and neutralinos (or gravitinos,  \ ...\ ).}
\label{sec:ssm}

\vspace{-1.5mm}

\subsection{Standard decays into neutrinos.}

\vspace{-1.5mm}

The expected $\Upsilon$ decay rate into {\it neutrinos\,} (\ref{upsnu0}), 
which involves the vector part in the weak neutral 
current $J_Z=$
$J_3-\sin^2\theta\ J_{\rm em}\,$ of the $b$ quark, is obtained from
\be
\label{upsnu}
\ba{c}
\dis
\!\!\frac{B(\Upsilon\to\nu\bar\nu)}{B(\Upsilon\!\to e^+e^-)}\, \simeq \,
\dis
\frac{3}{2} 
\left(\frac{\frac{1}{2}\  {(-\frac{1}{4}+\frac{1}{3}\sin^2\theta)}\,(g^2+g'^2)/m_Z^2}{-\,\frac{e^2}{3}/m_\Upsilon^2}\right)^{\!\!2} 
\vspace{2.5mm}\\
\simeq
\dis
\ \frac{27 \ G_F^2\,m_\Upsilon^4}{64\,\pi^2\,\alpha^2}\ \,
(-1+\frac{4}{3}\sin^2\theta)^2 \ \simeq\ 4\ 10^{-4}\ ,
\ea
\ee
with $B(\Upsilon\!\to e^+e^-)\simeq\,2.4\ \%\,$ or possibly $2.5\ \%$.
The smallness of the resulting $\,B(\Upsilon\to\nu\bar\nu)\,\simeq\,10^{-5}\,$
reflects the smallness of $\,(m_{\Upsilon}/m_Z)^4$ $\approx 10^{-4}$ \cite{fayetkaplan,chang}.

\vspace{2mm}

In a similar way,
\be
\ba{c}
\vspace{-7mm}\\
\dis
\!\frac{B(\psi\to\nu\bar\nu)}{B(\psi\to e^+e^-)} \, \simeq \,\frac{3}{2} \,
\left(\frac{\frac{1}{2}\  {(\frac{1}{4}-\frac{2}{3}\sin^2\theta)}\,(g^2+g'^2)/m_Z^2}{\frac{2\,e^2}{3}/m_\Upsilon^2}\right)^{\!\!2} 
\vspace{2.5mm}\\
\dis
\ \ \ \ \ \simeq\  \frac{27 \ G_F^2\,m_\psi^4}{256\,\pi^2\,\alpha^2}\ \
(1-\frac{8}{3}\sin^2\theta)^2 \ \simeq\ 4\ 10^{-7}\ ,
\vspace{-3mm}\\
\ea
\ee

\vspace{3mm}
\noindent
the precise value depending on how renormalisation effects are taken into account.
With  $B\,(\psi\!\to\! e^+e^-)\simeq\,6\ \%$ this
leads to a very small $\,B\,(\psi\to\nu\bar\nu)\,\simeq \,(2$ to $3)\ 10^{-8}$ \cite{fayetpsi,chang}, well below
present experimental sensitivity \cite{bes}.

\vspace{-2mm}

\subsection{\vspace{1.5mm}
Decays into light neutralinos, \hbox{in the MSSM or N(n)MSSM.}}

\vspace{-4mm}

\begin{figure}[htb]
$$
\epsfig{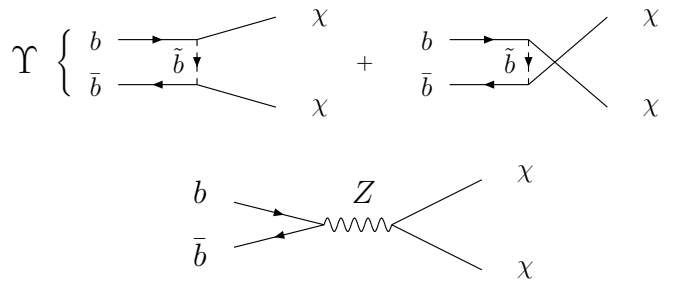}
$$
\caption{\em \small $\Upsilon$ decay into light neutralinos,
induced by squark or $Z$ exchanges (ignoring a possible $\,U$ contribution
as in Fig.\,\ref{fig:upschi}).
}
\label{fig:upsphot}

\end{figure}

The branching fraction for the pair-production of light {\it neutralinos\,} by squark 
or $Z$ exchanges (Fig.\,\ref{fig:upsphot}) is also expected to be small.
The amplitudes involving the exchanges of the two spin-0 squarks 
$\,\tilde b$ (Fig.\,\ref{fig:upsphot}) are induced in the local limit approximation by
\be
\bar b \,\gamma^\mu\, b \ \ \bar \chi \,\gamma_\mu\gamma_5 \chi \ \ 
\ee
4-fermion effective interactions proportional to $1/m_{\tilde b}^{2}$ \cite{plb79},
in a way which depends on the  composition (gaugino/ higgsino) of the neutralino $\chi$ considered, and of the mass matrix of the two squark
fields $\widetilde {b_L}$ and $\widetilde {b_R}\,$.
Indeed as $\Upsilon$ has the same quantum numbers $1^{--}\,$ as the photon, 
the $\,\bar b\,\gamma^\mu\gamma_5 \,b, \ \bar b\, b $ and $\bar b\gamma_5  b$ operators, which have $\,C = +\,$,  
\,cannot contribute to the decay, 
\,nor $\,\bar b\,\sigma^{\mu\nu}\,b\,$ (which has $C=-$) \,as  $\,\bar \chi \,\sigma_{\mu\nu}\chi\,$
vanishes identically for a Majorana $\chi$, \,as for $\,\bar\chi\,\gamma_\mu\chi\,$.

\vspace{1.5mm}

These amplitudes  may be easily compared with the $Z$ amplitudes for $\,\Upsilon\to\nu\bar\nu$.
They are very small for the two $\,\tilde b\,$  $\,\simge$ 100 GeV, corresponding to an invisible
branching fraction into $\,\chi\,\chi$ of less than $\,5\ 10^{-8}$  \cite{dreiner}
\footnote{In a similar way, $\,\tilde b$-exchange contributions 
to $\,\Upsilon\to\nu\bar\nu$ decay amplitudes in a $R_p$-violating theory \cite{barbier} are also very small \cite{chang}.}.
This leaves us with the $Z$ amplitude, to be discussed soon.

\vspace{1.5mm}

But what about possible contributions from the exchanges of neutral spin-0 particles\,?
The question arises especially as neutral particles such as a pseudoscalar $a\,$ or scalar $h\,$
(with 
$\,\bar b\,\gamma_5\,b\,$ or $\,\bar b\,b$ couplings to the $b$ quark)
could be light when the MSSM is extended to include a singlet $S$ with a $\,\lambda \ H_1 H_2 \,S\,$ trilinear superpotential term \cite{nmssm}, in connection with associated global $\,U(1)$ symmetries 
of 2-Higgs-doublet models, which may be almost-spontaneously broken \cite{dermisek,plb09}. 
Such light spin-0 particles might conceivably induce relatively large $\,\Upsilon \to\chi\,\chi$
invisible decay amplitudes.

\begin{figure}[htb]
$$
\epsfig{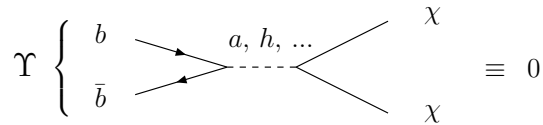}
$$
\caption{\em \small Invisible decay $\ \Upsilon\,\to\,\chi\,\chi\,$ that might be induced by \  $a\,$ or $\,h$, ...\,. 
These amplitudes vanish identically,
as for Dirac (\,$\Upsilon\to\,\chi\,\bar \chi$)
\,or spin-0 \,($\,\Upsilon \to \,\varphi\,\bar\varphi$) \,dark matter particles. The production of identical spin-0 particles,
$\,\Upsilon\to\varphi\,\varphi$, \,is forbidden by Bose statistics.
}
\label{fig:upsviascal}
\end{figure}

But $\,\Upsilon \to\chi\,\chi\,$ cannot proceed through the virtual production of a 
\hbox{spin-0} boson with pseudoscalar or scalar couplings to $\,\bar b \,\gamma_5\, b\,$ 
or $\,\bar b \,b$.
The  $\,b\,\bar b\,$ annihilation of the $\,\Upsilon$, a $C=-$ state, can only occur 
(independently of the fact that $C$ may be conserved or not in the decay)
through a $\,C=-$ hadronic operator such as 
the quark vector current $\,\bar b\,\gamma^\mu\,b$, \,but not the axial current $\,\bar b\,\gamma^\mu\gamma_5\,b$,
\,nor $\,\bar b\,\gamma_5\,b\,$ or $\,\bar b\,b$, 
\,which all have $\,C=+\,$. We thus always have

\vspace{-6mm}
\be
{\cal A} \ 
\left( \hbox{\LARGE \phantom{A}} \right. 
\!\!\!\!\Upsilon \stackrel{
\ba{c}
\hbox{\footnotesize spin-0}
\vspace{-.5mm}\\ 
\hbox{\footnotesize {$\  \ (a,\, h,\,...)\ \ $}}
\\
\ea
}{\hbox{\Large $\longrightarrow$}} \ \chi\ \chi
\left. \hbox{\LARGE \phantom{A}} \!\!\!\! \right)
\ \,= \,\ 0\ \ ,
\ee
as represented in Fig.\,{\ref{fig:upsviascal}.

\vspace{2mm}

One remains, in the MSSM as well as in the N or nMSSM, with the virtual-$Z$ amplitude for 
$\,\Upsilon\to\chi\,\chi\,$ (Fig.\,{\ref{fig:upsphot}). 
The $Z\,$ does not couple to neutral gauginos, only to higgsinos $\,\widetilde{h_1}\!\!\, ^0$ and $\widetilde{h_2}\! \,^0$, with opposite signs. With $\,\chi=$ $\alpha_1\, \widetilde{h_1}\!\!\, ^0 +$ $\,\alpha_2\,\widetilde{h_2}\! \,^0
+\,...\ $, \,the $Z$ coupling \,(written as a coupling to a chiral $\chi_L$)\, is $\,|\alpha_1|^2-|\alpha_2|^2$ times the $Z$ coupling to an ordinary neutrino $\nu_L$. This leads to a contribution to the invisible width of the $\Upsilon$ fixed by $\,(|\alpha_1|^2-|\alpha_2|^2)^2$.
Its size  is experimentally limited as the neutralino $\chi$ should not contribute too much to the invisible decay width of the $Z$. I.e. conservatively, considering \,``$N_\nu$''~$=\,2.92\,\pm\,.05$ from the direct measurement of the $Z$ invisible width, at most $\,\simeq 10\ \%$ 
of the contribution of a single neutrino flavor. As a result we get the estimate
\be
\hbox {\framebox [8.3cm]{\rule[-.6cm]{0cm}{1.45cm}$
\ba{ccc}
B\,(\Upsilon\to\chi\chi) &\simeq&
(|\alpha_1|^2-|\alpha_2|^2)^2\ \,\beta^3\  B\,(\Upsilon\to\nu_e\bar\nu_e)
\vspace{3mm}\\
&\simle&\ .1\ 
\beta^3\  B\,(\Upsilon\to\nu_e\bar\nu_e)\ \,
\simle \ 3\ 10^{-7}\ \ .
\ea
$}}
\ee
(The $\,\beta^3$ factor, with $\,\beta=v_\chi/c$,  is associated with the Majorana character of the neutralino $\chi$, coupled to the $Z$ through 
its axial current $\,\bar\chi\,\gamma^\mu\gamma_5\,\chi$.)

\vspace{2mm}

This analysis applies as well to invisible $\,\psi$ decays, leading to

\vspace{-5mm}
\be
B\,(\psi\to\chi\,\chi)\ \,\simle\ .1\ 
\beta^3\  B\,(\psi\to\nu_e\bar\nu_e)\ \,
\simle \ 10^{-9}\ \ ,
\ee 
assuming the two squarks $\,\tilde c\,$ to be somewhat heavier than $\,\simeq\,$ 200 GeV
so that their contribution is negligible \cite{dreiner}.

\subsection{\vspace{1mm} Spin-0 induced decays\,?}

\vspace{-2mm}

More generally, could exchanges of light neutral \hbox{spin-0} particles,
as represented in  Fig.\,\ref{fig:upsviascal}\, for neutralinos $\chi$, \,lead to significant 
invisible $\psi\,$ or $\,\Upsilon\,$ branching fractions, independently of the nature of the dark matter particle\,? 
As we saw invisible  $\,\psi$ or $\Upsilon$ decays
cannot be directly induced by   \hbox{spin-0} bosons 
coupled to $\,\bar b \,\gamma_5\, b\,$ or $\,\bar b \,b$,
since the  $b\,\bar b\,$ annihilation can only occur through a $\,C=-$ hadronic operator, so that
\be
{\cal A}\ (\,\psi/\Upsilon \! \stackrel{\ba{c}\hbox{\footnotesize spin-0 part.}\\ \ea }{\hbox{\Large $\longrightarrow$}} \!
 \hbox{invisible}\, )\ \,= \,\ 0\ \ .
\ee
This applies independently of the spin of the final particles, for Majorana or Dirac ($\,\Upsilon\to\chi\,\chi\,$ or $\,\chi\,\bar\chi\,$) as well as spin-0 ones ($\varphi\,\bar \varphi$).
The decay $\,\Upsilon\to \varphi \,\varphi\,$ into a pair of identical spin-0 particles, in a $\,L=1$ state, 
is in any case forbidden by angular momentum conservation and Bose statistics, independently of its possible mediator(s). In particular,
\be
B\,(\,\psi/\Upsilon\rightarrow \hbox{pair of self-conj.~spin-0 part.})\ \, \equiv\ \, 0\ \ . 
\ee

\subsection{Decays into gravitinos + neutralinos.}

\begin{figure}[tb]
$$
\epsfig{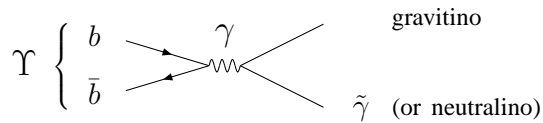}
$$
\caption{\em \small Decay $\,\Upsilon\to$ gravitino + photino (or neutralino) \cite{fayetpsi,prd06}.
The amplitude is $\,\propto\,\kappa\,e/m_{3/2}$. The photino or neutralino is unobserved if sufficiently light, otherwise it can decay into $\ \gamma$ + gravitino.}
\label{fig:upsgrav}
\end{figure}

A spin-$\frac{3}{2}$ gravitino, although coupled with gravitational strength ($\,\propto \kappa=\!\sqrt{\,8\pi ~G_N}
\simeq 4\ 10^{-19}$ GeV$^{-1}$),
could still interact ``weakly'' or even relatively ``strongly'' with standard model particles, if it is sufficiently light, much less than 1 eV \cite{plb79,grav}.
$\psi\,$ or $\Upsilon$ could then decay into a gravitino + a photino or neutralino
(the latter remaining also unobserved if it is light enough).
These decays are primarily induced by a virtual photon coupled to the $c$ or $b$ quark (Fig.\,{\ref{fig:upsgrav}),
having with the gravitino and photino a non-diagonal $q^2$-dependent charge-like effective coupling $\,\kappa\,q^2/(m_{3/2}\,\sqrt 6)=q^2/d\,$. It leads to an effective pointlike 4-fermion interaction, which could be large if the gravitino mass $m_{3/2}\,$ is really very small.
\vspace{.5mm}
The branching fractions are obtained in the photino case \vspace{1mm}
from 
\vspace{-1.2mm}
 {\large $\, (\,\frac{\kappa}{m_{3/2}\,\sqrt 6}\,/\,\frac{e}{m_{\psi/\Upsilon}^{\,2}}\,)^2$}, \ 
independently 
\vspace{1mm}
of the charge  2/3 or $\,-\,1/3$ of the annihilating $c$ or $b$ quark, 
leading to\,  \cite{fayetpsi,prd06}:
\be
B\,(\psi/\Upsilon \to  \hbox{\small gravitino}\, + \tilde\gamma) \,  \simeq \, 
\hbox{\small $\dis \frac{G_N\,m_{\psi/\Upsilon}^{\,4}}{3\ m_{3/2}^{\,2}\,\alpha}$}\ \
B\,(\psi/\Upsilon\to e^+e^-)\, .
\ee
If the photino $\tilde \gamma$ is replaced by a neutralino $\chi$, this expression should be multiplied by $\,\cos^2$ of the neutralino/photino mixing angle.

\vspace{2mm}

The resulting lower limit on the gravitino mass, now
\be
m_{3/2}\ >\  4.4\ 10^{-7}\ \hbox{eV}\ \ \hbox{ (with light quasistable\ \,$\tilde \gamma$)} \ \ ,
\ee
from $\Upsilon$ decays, is improved by $\,\sqrt{\, 5\ 10^{-2}/(3\ 10^{-4})}\,\simeq  13\,$ 
over the $\,3\ 10^{-8}$  eV
obtained from a $5\ \%$ CLEO limit in 1984 \cite{prd06,upsinv}.
We also get, from invisible $\psi$ decays \cite{bes}, 
$\,m_{3/2} >  4.8\ 10^{-8}$ eV, improved by
$\,\simeq \sqrt{\, 7\ 10^{-3}/(7.2\ 10^{-4})}$ $\,\simeq  3.1\,$ over the first $\,1.5\ 10^{-8}$  eV limit of \cite{fayetpsi}.
These limits, however, are largely superseded by those already obtained long ago at higher energies, 
e.g. in $\,e^+e^-$ annihilations \cite{plb86},
as the associated gravitino-neutralino production cross sections, increasing with energy proportionally to $s$, are easier to detect at higher energies.

\vspace{-1mm}

\section{\vspace{1.5mm}
Decays \hbox{into scalar dark matter\ } \hbox{through heavy-fermion exchanges}.}

\label{sec:scal}

\vspace{-1mm}

\subsection{Invisible decays \boldmath $\,\Upsilon\to\,\varphi\,\bar\varphi\,$.}

\vspace{-1mm}

Annihilations of spin-0 light dark matter particles may also occur through exchanges of new heavy fermions 
such as mirror fermions \cite{bf}. The dark matter particle $\,\varphi\,$ is a mixing of a singlet 
and neutral component of an electroweak triplet, so that it is not directly coupled to the $Z$ boson.
The decay 
\be
\Upsilon\ \ \stackrel{\ba{c}\hbox{\small $b_M$}\vspace{0mm}\\ \ea}{\longrightarrow}\   \underbrace{\ \varphi\ \bar\varphi\ }_{\hbox{\small \it invisible}}\  
\ee
could then be induced by the exchange of a new heavy quark $b_M$ (Fig.~\ref{fig:upssc}).  
The non-diagonal couplings of the spin-0 dark matter field $\,\varphi\,$ 
to the $b$ quark and its heavy (mirror) partner $\,b_M$ are given, in terms of the two chiral Yukawa coupling constants $\lambda_{bL}$ and $\,\lambda_{bR}$, by
\be
\label{yukawa}
{\cal L}=-\,m_{b_M}\ \overline {b_M\!}\ b_M\,+\, [\varphi^\dagger\  \overline {b_M\!}\ \ (\lambda_{bL}\,b_L+\lambda_{bR}\,b_R) +\, \hbox{h.\,c.}]\,+\,...\, .
\ee

\vspace{1mm}

The exchange of a heavy $\,b_M$ leads in the local limit approximation $m_{b_M}\gg m_ b$, 
in which
\be
b_M\ \ =\ \ \frac{1}{m_{b_M}\!}\ \ \varphi^\dagger\ (\lambda_{bL}\,b_L+\lambda_{bR}\,b_R)\ +\ ... \ \ ,
\ee
to a dimension-5 effective interaction between $b$ quarks and dark matter particles, given by
\be
\label{eff5}
{\cal L}_{\rm eff}^{\,5}\ =\ \ m_{b_M}\  \overline {b_M\!} \ b_M\ =\ \frac{\lambda_{bR}^{\,*}\,\lambda_{bL}}{m_{b_M}}\ \ \overline{b_R\!\!}\ \,b_L\ \varphi^\dagger\varphi\ +\ \hbox{h.c.}\ \  .
\ee
This operator, however, cannot contribute to invisible $\Upsilon$ decays,
$\,\bar b \, b$ and $\,\bar b \,\gamma_5\, b\,$ having  $\,C=+$, \,unlike $\,\Upsilon$ \cite{prd06}.
We are then interested, for next-order contributions to the amplitude, in the ($C=-$) quark current $\,\bar b\,\gamma^\mu\,b$ (or $\bar c\,\gamma^\mu c$), 
\,effectively coupled to the spin-0 dark matter current 
$\,\varphi^\dagger \, i\!
\hbox{\small $
\ba{c}\vspace{-7.8mm}\\ \leftrightarrow \vspace{-1.2mm}\\ \partial\vspace{-2mm}\\ \ea
$}
 \!\! _\mu \,\varphi$, \,proportionnally to $\,1/m_{b_M}^{\,2}$.

\vspace{2mm}

The invisible decay amplitude, vanishing at order $1/m_{b_M}$, \,is proportional to  $1/m_{b_M}^{\,2}$.
This contrasts with
$\,\varphi\,\bar \varphi\to f\bar f\,$ dark-matter annihilation amplitudes (and heavy-fermion contributions to lepton anomalous magnetic moments). These are proportional to
$\,1/m_{f_M}$  ($f_M$ being the exchanged fermion) in the non-chiral case where $\lambda_{fL}\,$ and $\,\lambda_{fR}$ couplings are both present, as seen from the formulas involving the fermion $\,f$, analogous to (\ref{yukawa},\ref{eff5}).

\begin{figure}[tb]

$$
\epsfig{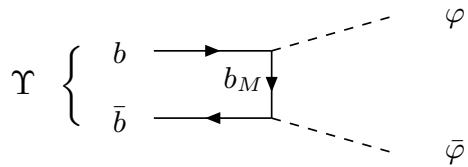}
$$
\caption{\em \small Invisible decay into a pair of (non self-conjugate) spin-0 dark matter particles, induced by a heavy quark $\,b_M$ \cite{bf}. The amplitude, $\,\propto 1/m_{b_M}^{\ 2}$,  is expected to be small.}
\label{fig:upssc}
\end{figure}

\vspace{2mm}

The invisible decay amplitude is thus usually expected to be very small, at most of the order of weak-interaction amplitudes 
for $\,m_{b_M} \simge 100 $ GeV and Yukawa couplings 
$\,\simle\,$  electroweak gauge couplings $g$ and $g'$.

\vspace{2mm}
 Even for 
$m_{q_M}$ as low as $ \approx 100$ GeV, we would need, to get from heavy $b_M$ exchanges a significant $\Upsilon\to \varphi\,\bar\varphi\,$ 
branching fraction possibly approaching the experimental limit (\ref{limups}),
rather large Yukawa couplings $\,\lambda_b$
of $\,\varphi\,$ to $b$ and  $b_M$, as compared to $g$ and $g'$, corresponding to a stronger-than-weak interaction at $\sqrt s\simeq m_\Upsilon$.
But analogous Yukawa couplings to charged leptons \,($\lambda_{l}$)\, or light quarks \,($\lambda_{q}$)\,
of similar size
would tend to lead to excessively large effects in $e^+e^-$ scatterings ($\,e^+e^-\to\gamma \ \varphi\,\bar\varphi\,$), \,anomalous magnetic moments of charged leptons,
with large contributions \cite{bf}
\be
\delta a_l \ \simeq\ \frac{\Re\,(\lambda_{lR}^{\,*}\,\lambda_{lL})}{16\,\pi^2}\ \frac{m_l}{m_{lM}}\ \ ,
\ee 
$\,K^+\to\,\pi^+\,\varphi\,\bar\varphi\,$, \,... .

\vspace{2mm}

The residual invisible $\Upsilon$ decay amplitude, proportional to $1/m_{bM}^{\,2}$, \,may be estimated
from the relevant part in the heavy $\,b_M$ propagator,
$\,q\!\!\!/\ /(q^2-m_{b_M}^{\,2})$, 
\,replacing the virtual momentum
$\,q=\frac{1}{2}\, (p_1-k_1) -\frac{1}{2}\,(p_2-k_2)\,$ 
by $\,\frac{1}{2}\,(k_2-k_1)$
\,(as the terms proportional to $\,p\!\!\!/_1\ (b)\,$ and $\,p\!\!\!/_2\ (\bar b)\,$ lead again to  $C$-even operators
which do not contribute).
This leads to the dimension-6 operator,
\be
\left(\,|\lambda_{bL}|^2\ \bar b_L\,\gamma^\mu\,b_L + |\lambda_{bR}|^2\ \bar b_R\,\gamma^\mu\,b_R\,\right)
\ \ \frac{\varphi^\dagger\  i \hbox{\small $
\ba{c}\vspace{-7.8mm}\\ \leftrightarrow \vspace{-1.2mm}\\ \partial\vspace{-2mm}\\ \ea
$}
 \!\! _\mu \,\varphi}{2\,m_{b_M}^{\,2}}\ \ ,
\ee
in which we retain the ($C$-odd) $\,\bar b\,\gamma^\mu b\,$ contribution 
\be
\frac{|\lambda_{bL}|^2+ |\lambda_{bR}|^2}{2}\ \ \frac{1}{2}\ \ \bar b\,\gamma^\mu\,b\
\ \ \frac{\varphi^\dagger\  i \hbox{\small $
\ba{c}\vspace{-7.8mm}\\ \leftrightarrow \vspace{-1.2mm}\\ \partial\vspace{-2mm}\\ \ea
$}
 \!\! _\mu \,\varphi}{m_{b_M}^{\,2}}\ \ .
\ee
Comparing with $\,\Upsilon\to e^+e^-$, \,given by
\be
\frac{e^2}{3}\ \ \bar b\,\gamma^\mu\, b\ \ \frac{\bar e\,\gamma_\mu\,e}{m_\Upsilon^{\,2}}\ \ ,
\ee
we get
\be
\label{rapport}
\ba{l}
\dis
\frac{B\,(\,\Upsilon\to\varphi\,\bar\varphi\,)}{B\,(\,\Upsilon\to e^+e^-\,)}\simeq 
\dis
\left(\, \frac{|\lambda_{bL}|^2+ |\lambda_{bR}|^2}{2}\ \frac{3}{2\,e^2} \,\right)^2 
\frac{m_\Upsilon^{\ 4}}{m_{b_M}^{\,4}}\ \,\frac{1}{4}\ \beta^3
\vspace{3mm} \\
\simeq \dis\ .536\ 10^{-2}\ 
\left(\, \frac{|\lambda_{bL}|^2+ |\lambda_{bR}|^2}{2} \,\right)^2 
\frac{(100\ \hbox{GeV})^4}{m_{b_M}^{\,4}}\ \ \beta^3\, .
\ea
\ee
$\frac{1}{4}\ \beta^3\,$ is associated with the pair-production of \hbox{spin-0} \ \ particles, 
\vspace{-1mm}
as compared to massless spin-$\frac{1}{2}$ Dirac ones, with
$
\,\beta=\frac{v_\varphi}{c}=\left(1-\frac{4\,m_\varphi^2}{m_\Upsilon^2}\right)^{\!1/2}$.

\vspace{2mm}

With leptonic branching fractions of $\Upsilon$ 
of about $(2.38 \pm .11)\, \%, \ (2.48 \pm .05)\, \% $ and $\,(2.60 \pm .10)\, \% $ \,for $e^+e^-,\ \mu^+\mu^-$ and $\tau^+\tau^-$, \,respectively, and choosing to use the value $\simeq 2.48 \ \%$ to evaluate the invisible branching fraction (\ref{rapport}), we get
\be
\label{fraction}
\framebox [8.5cm]{\rule[-1.cm]{0cm}{2.2cm}$
\ba{l}
\ \ \ B\,(\,\Upsilon\to\varphi\,\bar\varphi\,)\ \ \simeq
\vspace{2mm}\\
\ \ \ \ \dis \ 1.33\ 10^{-4}\ 
\left(\, \frac{|\lambda_{bL}|^2+ |\lambda_{bR}|^2}{2} \,\right)^2 \ 
\frac{(100\ \hbox{GeV})^4}{m_{b_M}^{\,4}}\ \ \beta^3.
\ea
$}
\ee

\vspace{2mm}

To get an upper bound on the Yukawa couplings, however, we may use $ \simeq 2.38 \ \%$ but 
as a lower bound on the branching fraction into $e^+e^-$, 
the 1.33 in (\ref{fraction}) getting replaced by about 1.275.
This leads to 
\be
\ba{l}
\dis \sqrt{\ \frac{|\lambda_{bL}|^2+ |\lambda_{bR}|^2}{2}\,}\ <\ 
\vspace{1mm}\\
\dis
\ \ \  9.4\ \ [\ \hbox{lim}\,(B_{\hbox{\footnotesize inv}}(\Upsilon))\ ]^{1/4}\ \frac{m_{b_M}}{100\ \hbox{GeV}}\  
\left(1-\frac{4\,m_\varphi^2}{m_\Upsilon^2}\right)^{-\,3/8}\!,
\ea
\ee
i.e., for  $B_{\hbox{\footnotesize inv}}(\Upsilon) <\ 3\ 10^{-4}\,$ \cite{babarinv}, 
\be
\dis \sqrt{\,\frac{|\lambda_{bL}|^2+ |\lambda_{bR}|^2}{2}}\ < 
\underbrace{1.24\  \left(1-\frac{4\,m_\varphi^2}{m_\Upsilon^2}\right)^{-\,3/8}}_{\hbox{ $<\  1.5$}\ \ \hbox{\footnotesize for}\, \hbox{ $m_\varphi< 3$}\ \hbox{\small GeV}}  \frac{m_{b_M}}{100\ \hbox{GeV}}\ \ ,
\ee
assuming it makes sense to consider such large values of the Yukawa couplings.
This upper bound is in agreement with the one obtained in \cite{ye} from a similar analysis of
$\,\Upsilon\, $ decays into scalar dark matter particles.

\vspace{2mm}
These invisible decays into scalar dark matter induced by heavy-fermion exchanges are likely to remain unaccessible,
due to the smaller expected values of the Yukawa couplings.
We may then turn to the radiative decays \cite{prd06}
\be
\Upsilon\ \ \stackrel{\hbox{\small$b_M$}}{\longrightarrow} \ \ \gamma\underbrace{\ \varphi\ \bar \varphi\ }_{\hbox{\small \it invisible}}\ \ ,
\ee
which may be induced by the effective dimension-5 operator (\ref{eff5}), with a branching fraction 
$\,\propto\,\alpha \  |\lambda_{bR}^*\,\lambda_{bL}|^2\ m_\Upsilon^{\,2}/$ $m_{b_M}^{\,2}$  (to be compared, for example, with a branching fraction into $\,\gamma\ \pi^+\pi^-$ of about $6\ 10^{-5}$).
We would need, again, rather large values of the Yukawa couplings for this branching fraction into $\,\gamma\, +$ 
{\it invisible\,} to be significant, as also discussed in \cite{ye}.

\vspace{-3mm}

\subsection{Invisible decays \boldmath $\,\psi\to\,\varphi\,\bar\varphi\,$.}

\vspace{-2mm}

For invisible $\,\psi\,$ decays we have, in a similar way,
\be
\ba{r}
\dis
\frac{B\,(\,\psi\to\varphi\,\bar\varphi\,)}{B\,(\,\psi\to e^+e^-\,)}\ \simeq \
\dis
\left(\, \frac{|\lambda_{cL}|^2+ |\lambda_{cR}|^2}{2}\ \frac{3}{4\,e^2} \,\right)^2 
\frac{m_\psi^{\ 4}}{m_{c_M}^{\,4}}\ \,\frac{1}{4}\ \beta^3
\vspace{3mm} \\
\simeq \dis\ 1.54\ 10^{-5}\ 
\left(\, \frac{|\lambda_{cL}|^2+ |\lambda_{cR}|^2}{2} \,\right)^2 
\frac{(100\ \hbox{GeV})^4}{m_{c_M}^{\,4}}\ \ \beta^3\ .
\ea
\ee
With $\,B\,(\psi\to e^+e^-)\ \simeq\ 5.94\ \%\,$ we obtain
\be
\hbox {\framebox [8.4cm]{\rule[-.82cm]{0cm}{1.9cm}$
\ba{l}
B\,(\,\psi\to\varphi\,\bar\varphi\,)\ \ \simeq
\vspace{2mm}\\
\ \ \ \ \dis \ .91\ 10^{-6}\ 
\left(\, \frac{|\lambda_{cL}|^2+ |\lambda_{cR}|^2}{2} \,\right)^2 \ 
\frac{(100\ \hbox{GeV})^4}{m_{c_M}^{\,4}}\ \ \beta^3\ , 
\ea
$}}
\ee
leading to
\be
\ba{l}
\dis \sqrt{\,\frac{|\lambda_{cL}|^2+ |\lambda_{cR}|^2}{2}}\ <
\vspace{2mm}\\
\ \ \ \  32\ \
[\ \hbox{lim}\,(B_{\hbox{\footnotesize inv}}(\psi))\ ]^{1/4}\ \ \left(1-\frac{4\,m_\varphi^2}{m_\psi^2}\right)^{-\,3/8}\,\frac{m_{c_M}}{100\ {\rm GeV}}\ \ .
\ea
\ee
From the experimental limit  $\,7.2\ 10^{-4}\,$ \cite{bes} we get
\be
\dis \sqrt{\,\frac{|\lambda_{cL}|^2+ |\lambda_{cR}|^2}{2}}\ <\ 5.3\ \  \left(1-\frac{4\,m_\varphi^2}{m_\Upsilon^2}\right)^{-\,3/8}\ \frac{m_{c_M}}{100\ \hbox{GeV}}\ \ .
\ee
This mainly indicates that quite large Yukawa couplings would be required to get from heavy-fermion exchanges a significant branching fraction of $\psi$ into scalar dark matter.

\section{\vspace{1mm} \boldmath $\,U$-induced decays of $\,\psi\,$ and $\,\Upsilon$.}
\label{sec:U}

\vspace{-2mm}

\subsection{\vspace{1mm} \boldmath The $\,U$ as mediator of a new interaction.}

\vspace{-3mm}

We now turn again to the situation, that we consider more promising, of production and annihilation reactions induced by the light spin-1 $\,U$ boson, as represented later in Figs.~\ref{fig:upschi} and \ref{fig:ann3}.

\begin{figure}[thb]
$$
\epsfig{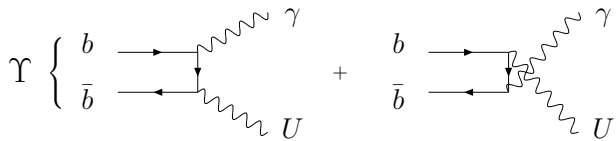}
$$
\caption{\em \small The decay $\Upsilon\to\gamma \,U$ \cite{U,plb09}, induced by the {\em axial\,} coupling 
$\,f_{bA}$ of the $b$ quark.
}
\label{fig:upsgammau}
\end{figure}

\vspace{2mm}

Meanwhile we recall that the $U$  may be directly produced in radiative $\psi\,$ and $\Upsilon$ decays, 
$\,\psi\to\,\gamma\,U\,$ and $\,\Upsilon\to\gamma\, U$, 
through its {\it axial} couplings to quarks $\,f_{cA}$ and $\,f_{bA}$, as shown in Fig.\,\ref{fig:upsgammau}.
The corresponding rates may be sufficiently small, a light $U$ behaving very much as an equivalent 
pseudoscalar, linear combination of doublet (interacting) and singlet (essentially ``inert'') components
\,\cite{U,step}. 
Upper limits on the production of a monochromatic photon + invisible particles
(or  a $\mu^+\mu^-$ pair, etc.), lead to strong upper limits 
on these axial couplings \cite{prd07,plb09}, e.g.~from the radiative production of invisibly decaying light $U$ bosons, 
\be
\left\{\ 
\ba{ccr}
|f_{bA}|& <& 4 \ 10^{-7}\ m_U (\hbox{MeV})/\sqrt{B_{\rm inv}(U)}\ \ ,
\vspace{2mm}\\
|f_{cA}|&<& 1.5 \ 10^{-6}\ m_U (\hbox{MeV})/\sqrt{B_{\rm inv}(U)}\ \ .
\ea \right.
\ee
The {\it \,vector\,} couplings $\,f_{cV}$ and $\,f_{bV}$, on the other hand,
may induce invisible $\,\psi\,$ and $\,\Upsilon$ decays into light dark matter particles  (Fig.\,\ref{fig:upschi}), possibly at a significant rate, as we now discuss.

\subsection{Invisible branching fractions.}

\vspace{-1mm}

If we take aside the possibility of large Yukawa couplings of scalar dark matter to the $b$ and $c$ quarks, discussed in Sec.~\ref{sec:scal},
invisible $\Upsilon$ and  $\psi$ decays 
only give significant new information on 
possible decays into light dark matter particles induced by a new (light) spin-1
$U$ boson with {\it vector} couplings to quarks, as illustrated in Fig.\,\ref{fig:upschi} \cite{fayetkaplan,bf,fermion,prd06}.
Decays into $\,\nu\bar\nu\,$ and $\,\chi\chi\,$ are induced by the vector parts in the $Z$ and $U$ currents, involving the same quark current $\,\bar b\,\gamma^\mu b\,$, \,or $\,\bar c\,\gamma^\mu c\,$.
\,The $\Upsilon$ branching fraction into light dark matter particles is given \cite{prd06}, for a spin-$\frac{1}{2}$ Majorana $\chi$ with an axial coupling $\,c_\chi/2\,$ to the $U$ boson (corresponding for a nearly massless $\chi$ to a coupling 
$c_\chi$ to $\,\chi_L$),
by
\be
\label{rapinvee}
\frac{B\,(\Upsilon\to\chi\,\chi)}{B\,(\Upsilon\to e^+e^-)}\ \simeq \
\dis
\frac{1}{2}\ \beta^3\  \left(\frac{c_\chi\,f_{bV}}{e^2/3}\right)^2\ 
 \frac{1}{\left( 1-\frac{m_U^{\,2}}{m_\Upsilon^{\ 2}}\right)^2}\ \ .
\ee

\vspace{-2mm}

\begin{figure}[h]
$$
\epsfig{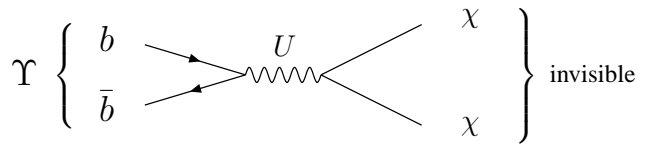}
$$
\caption{\em \small $\Upsilon$ decay into light dark matter particles, 
\,induced by a $\,U$ boson \cite{U} with {\em \,vector} couplings to $b$ 
\cite{fayetkaplan,prd06}.
This applies to spin-0\, or spin-$\frac{1}{2}$ Majorana or Dirac particles \cite{bf,fermion}, $\,\Upsilon\to \varphi\bar\varphi$, $\chi\chi\,$  or $\,\chi\bar\chi$.
The amplitude is $\propto\,c_\chi\,f_{bV}/(m_\Upsilon^{\ 2}-m_U^{\,2})$, compared to $\,e^2/(3\,m_\Upsilon^2)$ \,for $ \,\Upsilon\to e^+e^-$.
}
\label{fig:upschi}

\end{figure}

With 
$
\,\beta=\frac{v_\chi}{c}=\left(1-\frac{4\,m_\chi^2}{m_\Upsilon^2}\right)^{\!1/2}\!,
$
\,and using as before 
$
B\,(\Upsilon\!\to e^+e^-)\,\simeq \,2.48\ \% $,
\,we get in the Majorana case

\vspace{-5mm}
\be
\label{BR}
\dis
B\,(\Upsilon\to\chi\,\chi)\ \simeq\ 13.3\ 
\dis
\left(c_\chi\,f_{bV}\right)^2\ 
\hbox{\large $\dis 
 \frac{\left(1-\frac{4\,m_\chi^2}{m_\Upsilon^2}\right)^{3/2}\!}{\left( 1-\frac{m_U^{\,2}}{m_\Upsilon^{\ 2}}\right)^2\!}
$}
\ \ ,
\ee

\vspace{-5mm}
\noindent
i.e. 
\be
B\,(\Upsilon\to\chi\,\chi)\ \simeq\ 13.3\ \ (c_\chi\,f_{bV})^2\ \ [\,\beta^3\,] \  
\ee
for a light $U$ compared to $m_\Upsilon$. This allows in principle for relatively large 
(or on the other hand very small !) values of the invisible branching fraction into LDM particles. 
This is also compatible with an appropriate value of the annihilation cross section of these 
light dark matter particles \cite{fermion}, as discussed more in Secs.~\ref{sec:comp} and \ref{sec:rel}.
We get, for $m_U$ and $\,2m_\chi$ small compared to $\,m_\Upsilon$,
\be
\label{limex}
\hbox {\framebox [7.6cm]{\rule[-.47cm]{0cm}{1.2cm}$
\dis B\,(\Upsilon\to\chi\,\chi)\ \simeq\ \
1.33\ 10^{-3} \ \left(\,\frac{c_\chi\,f_{cV}}{10^{-2}}\,\right)^2 \  \ ,
$}}
\ee
$|c_\chi\,f_{bV}| \simeq 10^{-2}$, in particular, being now excluded by (\ref{limups}).

\vspace{2mm}

For invisible $\,\psi\,$ decays we get
\be
\label{bpsi}
\frac{B\,(\psi\to\chi\,\chi)}{B\,(\psi\to e^+e^-)}\ \simeq \
\dis
\frac{1}{2}\ \beta^3\  \left(\frac{c_\chi\,f_{bV}}{2\,e^2/3}\right)^2\ 
 \frac{1}{\left( 1-\frac{m_U^{\,2}}{m_\psi^{\ 2}}\right)^2}\ \ ,
\ee

\vspace{-1mm}
\noindent
leading  for a light $U$, with a $\,\Upsilon\,$ branching fraction into $e^+e^-$ of $\,(5.94 \pm 0.06) \ \%$, to
\be
\label{bpsi2}
\dis
B\,(\psi\to\chi\,\chi)\ \simeq\ 8\ 
\dis
\left(c_\chi\,f_{bV}\right)^2\ [\ \beta\,]^3
\ \ ,
\ee
i.e. 
\be
\label{bpsi3}
\hbox {\framebox [7.2cm]{\rule[-.47cm]{0cm}{1.2cm}$
\dis
B\,(\psi\to\chi\,\chi)\ \simeq\ \
8\ 10^{-4} \ \left(\,\frac{c_\chi\,f_{cV}}{10^{-2}}\,\right)^2 \  \ ,
$}}
\ee
for $m_U$ and $\,2m_\chi$ small compared to $\,m_\psi$.

\vspace{2mm}

This also applies to the pair-pro\-duction of non-self-conjugate spin-0 or spin-$\frac{1}{2}\,$ particles, with the replacements in (\ref{rapinvee}) \cite{prd06}

\vspace{-4mm}
\be
\label{subst}
\frac{1}{2}\ \ \beta^3\ _{\rm Majorana}\ \ \rightarrow\ \ 
\left\{\ \ba{ccc}
 \frac{1}{4}\ \beta^3\ &:& \hbox{spin-0,}
\vspace{2mm}\\
\beta^3 &:& \hbox{Dirac, axial,}
\vspace{2mm}\\
\frac{3}{2}\,\beta-\frac{1}{2}\,\beta^3  &:& \hbox{Dirac, vector,}
\ea \right.
\ee

\vspace{2mm}
\noindent
so that (\ref{limex},\ref{bpsi3}) are replaced, for light spin-0 particles, by
\be
\label{limdirac}
\left\{\ 
\ba{c}
\dis B\,(\Upsilon\to\varphi\,\bar\varphi)\ \simeq\ 
6.6\ 10^{-4} 
\left(\,\frac{c_\chi\,f_{bV}}{10^{-2}}\,\right)^2\ ,
\vspace{3mm}\\
\dis B\,(\psi\to\varphi\,\bar\varphi)\ \simeq\ 
4\ 10^{-4} \left(\,\frac{c_\chi\,f_{bV}}{10^{-2}}\,\right)^2\ ,
\ea \right.
\ee
and 4 times as much, for light spin-$\frac{1}{2}$ Dirac particles.

\subsection{Limits on \boldmath $\,U$ couplings}

\vspace{-1mm}

The $\,\Upsilon$ branching fraction  (\ref{BR}) (but now estimated with an electronic branching
fraction of $\,2.38\,\%$ to get an upper limit, which leads to replace 13.3 by 12.7) leads to 
\be
\label{formulelim}
\ba{l}
\dis|\, c_\chi\,f_{bV}| < \,.28\ \sqrt{\,\hbox{lim}\,(B_{\rm inv})}\ 
\left|\,1-\frac{m_U^{\,2}}{m_\Upsilon^{\ 2}}\,\right| \left(1-\frac{4m_\chi^2}{m_\Upsilon^2}\right)^{\!-3/4}\!\! ,
\ea
\ee
which simplifies into
\be
\label{formulelim2}
|\,c_\chi\,f_{bV}|  < \,\frac{e^2\,\sqrt 2}{3}\ \
 \left(\,\frac{\hbox{lim}\,(B_{\rm inv})}{B_{ee}}\,\right)^{\frac{1}{2}}\ \simeq \ .28\ \sqrt{\,\hbox{lim}\,(B_{\rm inv})}
\ee
for $\chi$ and $\,U$ light compared to $\Upsilon$.
From the B{\small A}B{\small A}R limit $\,3\ 10^{-4}$ \cite{babarinv} we get
\be
\label{maj}
\hbox {\framebox [8,1cm]{\rule[-.35cm]{0cm}{.85cm}$
\ |\,c_\chi\,f_{bV}| \ < \ \ 5\ 10^{-3}\ \ \ \ \hbox{(for $m_U,\, 2\,m_\chi \,\simle \, 2$ GeV)}\ ,
$}}
\ee
still approximately valid for $\,2\,m_\chi\,$ and $\,m_U$ smaller than $\,\simeq\,m_\Upsilon/2\,$
(and which may even be used up to $m_U \simeq 13$ GeV although it overestimates the limit for $m_U$ in the vicinity of $\,m_\Upsilon$).
As in the very light gravitino case this bound is improved by $\sqrt{\,5\ 10^{-2}/(3\ 10^{-4})} \,\simeq $ $13$, 
as compared to 
$|\,c_\chi\,f_{bV}| \simle\, 6\ 10^{-2}$ \cite{prd06} derived from the \,5\,\% CLEO limit on $B_{\rm inv}$.

\vspace{2mm}

It may be compared with the similar limit for the vector coupling of $\,c$,
\be
\label{majc}
|\,c_\chi\,f_{cV}| \ < \ \ .95\ 10^{-2}\ \ ,
\ee
still approximately valid for $m_U$ and $\,2\,m_\chi$ smaller than $\,\simeq \,m_\psi/2$,
deduced (cf.~(\ref{bpsi3})) from the recent BES II limit on invisible $\psi$ decays,
$\,B(\psi\to\chi\chi)\,<\, 7.2\ 10^{-4}$ \cite{prd06,bes}, which should be improved soon at BES III \cite{bes3}.

\vspace{2mm}
With the kinematic factors reestablished as in (\ref{formulelim}), 
these limits read, in the Majorana case,
\be
\label{maj2}
\dis |\,c_\chi\,f_{bV}| \ < \ 5\ 10^{-3}\  \left|\,1-\frac{m_U^{\,2}}{m_\Upsilon^{\ 2}}\,\right|\  \left(1-\frac{4m_\chi^2}{m_\Upsilon^2}\right)^{\!-3/4}\,,
\ee
and similarly for (\ref{majc}).
They disappear if $\,m_\chi$ approaches $\,m_\Upsilon/2\,$ (or $m_\psi/2$),
and get stronger if  $\,m_U$ approaches $\,m_\Upsilon\,$ (or $m_\psi$) as the cross sections get enhanced from the 
$\,U$ propagator effect.

\vspace{2mm}

The limits (\ref{maj},\ref{majc}), multiplied or divided by $\sqrt 2$ owing to (\ref{subst}),
apply to non-self-conjugate spin-0 or $\frac{1}{2}$ dark matter particles:
\be
\label{0dirac}
\ba{c}
\left\{\ 
\ba{ccc}
|\,c_\varphi\,f_{bV}| \ < \ \ \ 7\ 10^{-3} &:&\hbox{spin-0,}
\vspace{2mm}\\
|\,c_\chi\,f_{bV}| \ < \ 3.5\ 10^{-3} &:& \ \ \hbox{spin-$\frac{1}{2}$ Dirac;}
\ea \right.
\vspace{4mm}\\
\left\{\ 
\ba{ccc}
|\,c_\varphi\,f_{cV}| \ < \ 1.4\ 10^{-2} &:&\hbox{spin-0,}
\vspace{2mm}\\
|\,c_\chi\,f_{cV}| \ < \ \ \ 7\ 10^{-3} &:& \ \ \hbox{spin-$\frac{1}{2}$ Dirac.}
\ea \right.
\ea
\ee
The effects of $\,m_U,\,m_\chi$ are taken into account by reintroducing  $\left|\,1-\frac{m_U^{\,2}}{m_\Upsilon^{\ 2}}\,\right|\,\beta^{-3/2}\,$ as in (\ref{maj2}), 
\ or $\left|\,1-\frac{m_U^{\,2}}{m_\Upsilon^{\ 2}}\,\right|$ \
$(\frac{3}{2}\,\beta-\frac{1}{2}\,\beta^3)^{-1/2}\,$
for a vectorially-coupled Dirac particle.

\section{\vspace{1mm}
Compatibility with \hbox{relic abundance requirements.}}
\label{sec:comp}
\vspace{-1mm}

\begin{figure}[tb]
$$
\epsfig{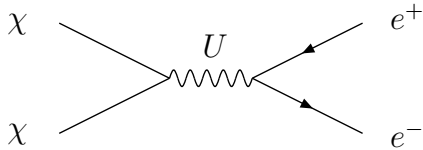}
$$
\caption{\em \small Dark matter annihilation into $e^+e^-$
induced by the $U$ boson \cite{bf,fermion,prd07}. 
This process should be stronger than for weak interactions, for a correct relic density of light dark matter.
}
\label{fig:ann3}
\end{figure}

How do the new limits (\ref{maj}-\ref{0dirac}) on $\,|c_\chi\,f_{bV}|\,$, $\,|c_\chi\,f_{cV}|$, etc., compare with the requirement that LDM particles must have sufficiently large annihilation cross sections at freeze-out time, 
for the correct relic abundance\,?
The cross section for ($P$-wave) annihilation of 
Majorana particles,
$\chi\,\chi\to e^+e^-$, \,may be written as \cite{fermion}
\be
\sigma_{\rm ann}\,v_{\rm rel} \ \simeq\ 
\hbox{ $ \dis
\frac{v_\chi^{\,2}}{.16}\ \left(\frac{c_\chi\,f_e}{10^{-6}}\right)^2
\ \left(\frac{m_\chi \times 1.8\ \hbox{\small MeV}}{m_U^{\,2}-4\,m_\chi^{\,2}}\right)^2
$}\ \ (4\ \hbox{pb})
\ee
disregarding $m_e$ for simplicity.
$c_\chi/2\,$ is the axial $\,U$ coupling to  $\,\chi$, 
$\,f_{eV}$ and $f_{eA}$ the  vector and axial couplings to the electron, with 
$\,f_e^{\,2}=f_{eV}^{\,2}+f_{eA}^{\,2}$. 
Similar expressions apply to the $P$-wave annihilations of non-self-conjugate
spin-0 particles, or spin-$\frac{1}{2}$ Dirac ones axially coupled to $U$.
The annihilation cross section of spin-$\frac{1}{2}$ Dirac particles 
with a vector coupling to $\,U$, however, can proceed through $S$-wave
and no longer involves the $\,v_\chi^{\,2}$ factor.

\vspace{2mm}

The ther\-mally-averaged total annihilation cross section at freeze-out time
should be 
$\,<\,\sigma_{\rm ann}\,v_{\rm rel}(\chi\chi\to f\bar f)\,>$ 
$\,\approx\,$ 4 - 5 pb, for a Majorana $\chi$ lighter than $\approx$ 1 GeV with $P$-wave annihilation. 
It corresponds to a cross section into $\,e^+e^- \,\approx\,$ 4 - 5 pb  times the annihilation branching fraction $B_{\rm ann}^{ee}$.
This requires \cite{fermion}
\footnote{$|c_\chi \,f_e|$ is slightly smaller for a heavier $m_\chi\approx $ a few GeV's, 
\\ 
with a slightly smaller required annihilation cross section. 
It is larger by $\sqrt 2\,$ for non-self-conjugate 
spin-0 or $\frac{1}{2}$ dark matter particles, with $P$-wave annihilation. It is
smaller (by a factor $\approx$ 2 to 3 fixed  by $1/v_\chi$ at freeze-out time) 
for $S$-wave annihilations of a vectorially-coupled 
Dirac $\chi$, as the required cross section ($<\sigma_{\rm ann}\,v_{rel}>\  \approx\,$ 4 pb) no longer includes 
the small $v_\chi^2$ factor \cite{fermion}.}:
\be
\label{sizecf}
\hbox {\framebox [8,1cm]{\rule[-.45cm]{0cm}{1.05cm}$
\dis
|\,c_\chi \,f_e\,|\ \,\simeq\ \,\left(B_{\rm ann}^{ee}\right)^{\frac{1}{2}}\  10^{-3}\ \
\frac{|m_U^{\,2}-4\,m_\chi^{\,2}|}{m_\chi\ (1.8\ \hbox{ GeV})}\ \ .
$}}
\ee

\vspace{2mm}

This is compatible with the constraint (\ref{maj},\ref{maj2}) from invisible $\Upsilon$ decays, 
for similar values of $f_{bV}$ and  $f_{e}$,
\,provided $\,m_U\,$ is not too large compared to
 $\sqrt {\,2\,m_\chi\times 1\ \hbox{GeV}}$. This also shows that a light $U$ is especially required for the smaller values of $\,m_\chi$, down to MeV scale.

\vspace{2mm}
For example with $m_U= 10$ MeV and $m_\chi \simeq 4$ (or 6) MeV 
as considered in \cite{fermion,prd07}, 
(\ref{sizecf})  would give
\be
|\,c_\chi \,f_e\,|\ \approx\ \, 3\ \ 10^{-6}\  ,
\ee
depending on $\,B^{ee}_{\rm ann}$ (here taken $\approx$ 40\ \%).
For a heavier $U$ we may get larger couplings, e.g.
\be
\hbox{up to}\ \  |\,c_\chi \,f_e\,|\ \simeq\ \frac{10^{-2}}{2\,m_\chi\,\hbox{(MeV)}}\ \ \ 
\hbox{for $m_U= 100$ MeV}\ \ 
\ee
(with   $B^{ee}_{\rm ann} \simeq 1$).
These values \cite{prd07} are generally  smaller than the 
new $\,5\ 10^{-3}$ upper limit (\ref{maj}) 
on
$|\,c_\chi \,f_{bV}\,|$.

\section{\vspace{1.2mm}
Estimating \boldmath $\psi$ and $\Upsilon$ invisible decays
\vspace{1.2mm}
from dark matter annihilation cross sections \hbox{\large ?}}
\label{sec:rel}

What about attempting to estimate the $\Upsilon$ invisible bran\-ching fraction into dark matter particles 
in a model-independent way from the dark matter annihilation cross section, 
with approximate expectations predicted $\ \simeq\, .41$~\%  
or \,.6 \%, or $\sim 1.8\ 10^{-3}$ for $P$-wave annihilation 
\cite{upsinv2,cleo2,mcelrath,kim}\,?
As we saw in the previous Sections, one cannot establish such predictions 
without taking into account, more specifically, essential features of these processes.
In fact we are dealing, not simply with $\,b\,\bar b\to\chi\chi$ and the inverse reaction
$\,\chi\chi\to b\,\bar b$, but with {\it different reactions}, and {\it at different energies}:
\be
\left\{\ 
\ba{cclcl}
b\ \bar b &\to& \ \ \chi\,\chi &&\hbox{at}\ \ \sqrt s=m_\Upsilon\ \ ,
\vspace{2mm}\\
\chi\,\chi &\to& l\,\bar l\,,\ \,q\,\bar q\ \ (q\neq b) & &\hbox{at a lower }\, \sqrt s \ \simeq \,2\,m_\chi\ .
\ea\right.
\ee

\vspace{1mm}
Furthermore a $U$ axially coupled to quarks would not contribute to invisible $\psi$ and $\Upsilon$ decays, 
\be
B(\psi\to\chi\chi)\ =\ B(\Upsilon\to\chi\chi)\ =\  0\ \ ,
\ee
while still inducing the desired dark matter annihilations into lighter quarks or leptons,
in the early Universe.
Direct spin-0 exchanges also do not contribute to these invisible $\psi$ and $\Upsilon$ decays.
For \hbox{spin-0} LDM particles $\varphi$ interacting non-chirally though heavy-fermion exchanges, annihilation cross sections 
are $\,\propto 1/m_{b_M}^{\,2}$~\cite{bf,fermion} \,but
$\,B(\Upsilon\to\varphi\bar\varphi)$ vanishes at this order \cite{prd06}. 
This branching fraction, $\,\propto (|\lambda_{bL}|^2+|\lambda_{bR}|^2)^2/m_{b_M}^{\,4}$ as we saw in Sec.\,\ref{sec:scal}, cannot be expressed proportionally to annihilation cross sections for $\,\chi\,\chi\to f\,\bar f$, 
$\,\propto |\lambda_{fR}^{\,*}\lambda_{fL}|^2 /m_{b_M}^{\,2}$ ...

\vspace{1mm}
It is thus essential to take into account the nature of the underlying process responsible 
for invisible decays and LDM annihilations, $\,U$ exchanges or possible heavy-fermion exchanges, ... ,
and treat correctly the fact that they occur {\it at different energies}, 
with in general {\it energy-dependent cross sections}.

\vspace{1mm}
Let us now concentrate on $U$-induced reactions.
Relations between $\,B_{\rm inv} (\Upsilon)$ (or $\psi$)\, and annihilation cross sections are implicit from the comparison between the upper limit (\ref{maj2}) on $\,|c_\chi\,f_{bV}|\,$ from  $\,B_{\rm inv}(\Upsilon)$, and 
the value (\ref{sizecf}) of $\,|c_\chi\,f_e|\,$ required for the correct relic density.
Taking $\,|\,c_\chi \,f_{bV}\,| \approx\,|\,c_\chi \,f_e\,|$ as seems natural 
in the absence of more specific informations on the nature of the couplings, we can write
the invisible branching fraction (\ref{BR}) proportionally to the annihilation cross section at freeze-out time,

\vspace{-7mm}
\bea
\label{uchi}
\!\!
\ba{ccl}
B\,(\Upsilon\to \chi\,\chi)&\propto& \!
\dis
(c_\chi \,f_{bV})^2\ 
 \left(\frac{m_\Upsilon^{\,2}}{m_\Upsilon^{\ 2}-m_U^{\,2}}\right)^2 
\vspace{2mm}\\
&\propto&
\
\dis
\frac{m_\Upsilon^{\ 2}}{4\,m_\chi^{\,2}}\, \left(\,\frac{4\,m_\chi^{\,2}-m_U^{\,2}}{m_\Upsilon^{\ 2}-m_U^{\,2}}\,\right)^2 
 ... \  <\,\sigma\,v >_{\hbox{\scriptsize \it  FO}}
\ea
\nonumber
\eea

\vspace*{-5mm}
\be
\ba{ccl}
&&
\hspace{-.5cm}
\propto \
\left\{\ \ba{ll}
\dis
\ \ \ \ \frac{m_\Upsilon^{\ 2}}{4\,m_\chi^{\,2}}\ \ \ (\,...\,)& 
\ \ \ 
\hbox{for  large }\,  m_U\, \simge\,2\,m_\Upsilon\ \ ,
\vspace{2mm}\\
\dis
\frac{m_U^{\,4}}{4\,m_\chi^{\,2}\ m_\Upsilon^{\ 2}}\ \ \ (\,...\,)
& 
\ \ \ 
\hbox{for}\ \, 4\,m_\chi\, \simle\,m_U\,\simle\,m_\Upsilon/2\ ,
\ea \right.
\ea
\ee

\vspace{1mm}

\noindent
where the \ ...\, include the extra factor $\,(f_{bV}/f_e)^2\ B_{\rm ann}^{ee}\,$.

\vspace{1mm}

The factor $\,m_\Upsilon^{\ 2}/(4\,m_\chi^{\,2})\,$ for a heavy $U$
originates from production and annihilation cross sections
growing with energy $\,\propto s$, from $\,\sqrt s \simeq 2\,m_\chi$ to $\,m_\Upsilon$, just as for ordinary weak interactions.
A stronger-than-weak annihilation cross section\ at $\,2\,m_\chi$, into $\,e^+e^-$ for example, corresponds to an effective Fermi-like coupling
$G'$ larger than $G_F$, that would then lead to stronger-than-weak processes at higher energies,
e.g. in $\,e^+e^-\to $ $\gamma\ \chi\,\chi$, \,which have not been observed \cite{fayetkaplan}.
Additional constraints are obtained from lepton anomalous magnetic moments, neutrino scatterings, parity-violation effects in atomic physics, ... \cite{prd07}.

\vspace{1mm}

The amplitudes corresponding to the exchanges of a {\it light} $\,U$, on the other hand, 
change behavior and start decreasing at energies
(or momentum transfer) larger than $\,\approx m_U$. Above this value the Fermi-like couplings $G'$, relevant in the local limit approximation corresponding to a heavy $U$, 
\vspace{-1mm}
\,have to be replaced by
\be
\label{G'}
G'\ \to\ \,G'\ \frac{m_U^2}{m_U^2-q^2}\ \simeq \ \,G'\ \frac{m_U^2}{-q^2} \ \ll G'\ \ \hbox{for a light}\  U\ . \vspace{-.5mm}\\
\ee
This mechanism was proposed to make the neutral current effects associated with
the exchanges of a new neutral gauge boson sufficiently small, when this one is light 
compared to the energies (or momentum transfer) considered \cite{U}.

\vspace{1mm}

For a $\,U$ somewhat lighter than $\Upsilon$, the $\Upsilon$ invisible fraction gets indeed inhibited as shown by (\ref{uchi},\ref{G'}), so that
\be
\label{relbinv}
B(\Upsilon\!\to \chi\chi) \propto (c_\chi f_{bV})^2 \approx (c_\chi f_e)^2 ... \propto
\frac{(m_U^{\ 2}-4\,m_\chi^{\,2})^2}{4\,m_\chi^{\,2}\,m_\Upsilon^{\,2}}\,   ...\, .
\ee
where the last \ ...\, include the factor $\,(f_{bV}/f_e)^2\ B_{\rm ann}^{ee}\,$.
This reflects again that $U$-induced cross sections grow like $s$ 
from $\,\simeq 4\,m_\chi^{\,2}\,$ but now only up to  $\,\approx m_U^{\,2}\,$, 
then decrease like $\,1/s\,$ between $\,m_U^{\,2}\,$ and $\,m_\Upsilon^{\,2}$, leading to a smaller result for
the invisible branching fraction. The lighter the $\,U$ (within the limits indicated as compared to $\,2\,m_\chi$), 
the easier it is to satisfy strong bounds from invisible $\Upsilon$ and $\psi$ decays, in particular.

\vspace{1mm}

This illustrates that there is no general way to predict
the invisible $\Upsilon\to\chi\chi$ branching fraction simply from the annihilation cross section of 
light dark matter particles at freeze-out time.

\vspace{-1.5mm}

\section{Non-perturbative effects.}

\label{sec:npert}

\vspace{-2mm}

These results could be affected by
non-perturbative exchanges of light $U$ bosons,
although this is not the situation we generally have in mind, as we would prefer the theory to remain perturbative \cite{prd07}.
For relatively large values of the $U$ coupling to dark matter particles (naively 
for $\alpha_\chi\!=c_\chi^{\,2}/4\pi\simge 1$, or similarly for $\,\alpha_\varphi$),
potentially large non-perturbative effects may have to be considered.
One could get an enhancement of the $\Upsilon\to\chi\chi$ branching fraction from $U$-mediated 
$\chi\chi$ interactions in the final state, 
especially if $2\,m_\chi$ or $2\,m_\varphi$ is relatively close to $\,m_\Upsilon$,
or one should consider $U$-radiative decays like $\Upsilon\to\chi\chi\,U$
(cf.~Figs.\,\ref{fig:upschiU} and \ref{fig:upschiU2}).

\vspace*{-1mm}

\begin{figure}[hb]
$$
\epsfig{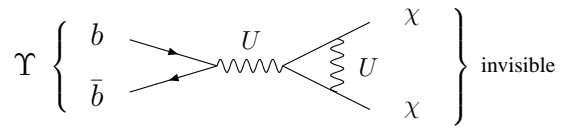} \vspace{-3mm}
$$
\caption{\em \small 
Radiative correction to $\,\Upsilon \to \chi\chi$, 
from $U$-mediated $\,\chi\,\chi\,$ interactions in the final state.
}
\label{fig:upschiU}
\end{figure}

\vspace*{-6mm}

\begin{figure}[hb]
$$
\epsfig{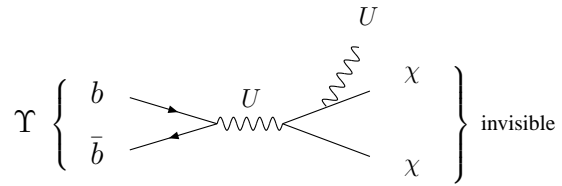} \vspace{-3mm}
$$
\caption{\em \small 
Radiative correction to $\,\Upsilon \to \chi\chi$, with emission of 
a light $\,U$ in the final state. 
}
\label{fig:upschiU2}
\end{figure}

For a spin-$\frac{1}{2}$ dark matter particle one has to take into account that a $U$ lighter than 
$\chi$ could effectively behave as an almost equivalent pseudoscalar $a$, with an effective coupling to $\chi$

\vspace{-7mm}
\be
\label{eq}
c_{\chi\,p}\ =\ c_\chi\ \ \frac{2\,m_\chi}{m_U}\ \ ,
\ee
leading to an enhanced effective coupling
$\,\alpha_{\chi\,p}=\alpha_\chi$ $(4\,m_\chi^2/m_U^2)\,$ \footnote{For a Majorana dark matter 
particle $\chi$, $c_\chi$ is defined as the coupling of $U^\mu$ to $\,\frac{1}{2}\ \bar\chi\,\gamma_\mu\gamma_5\,\chi$, and $c_{\chi\,p}$ as the corresponding coupling of 
the ``equivalent'' pseudoscalar $a$ to $\,\frac{1}{2}\ \bar \chi\,\gamma_5\,\chi\,$.} \cite{U,plb09}. 
The relative correction to the $\Upsilon\to\chi\chi$  amplitude (Fig.~{\ref{fig:upschiU})
or relative branching ratio into $\,U\chi\chi$ (Fig.~{\ref{fig:upschiU2}) 
are then proportional to
\be
\frac{1}{\pi}\ \alpha_{\chi\,p}\ =\ \frac{1}{\pi}\ \alpha_\chi\ \,\frac{4\,m_\chi^2}{m_U^2}\ =\ 
\frac{c_\chi^2}{\pi^2}\ \frac{m_\chi^2}{m_U^2}\ \ ,
\ee
which tells us if we are in the non-perturbative regime or not.
But $m_U<m_\chi$, or $c_\chi \,m_\chi$, does not correspond to 
\linebreak  
the situations 
we usually have in mind \cite{prd07},
and $m_U< $ $m_\chi$ 
would also lead to significant dark matter annihilations into $\,UU$ rather than $\,e^+e^-\!$, which may be disfavored\,\cite{nuss}.

\vspace{2mm}
Non-perturbative effects could also affect annihilation cross sections, possibly leading to a 
Sommerfeld-type factor \cite{somm}
enhancing annihilation cross sections of low-velocity dark matter particles.
Exchanges of light $U$'s between spin-$\frac{1}{2}$ dark matter particles 
would lead in the low-mass limit, in place of the usual Coulomb or Yukawa interactions, 
to a long-ranged spin-spin interaction
\be
\label{spsp}
c_\chi^{\,2}\ \ \frac{\vec\sigma_1 .\vec\sigma_2- 3\,\vec\sigma_1.\hat r\ \vec\sigma_2.\hat r}{4\,\pi\,m_U^{\,2}\ r^3}\ \ ,
\ee
as for ordinary fermions with axial 
couplings to the $U$~\cite{step,plb86force}.
The potential is the same as for the exchange of a quasimassless pseudoscalar $a$
with an effective coupling given by (\ref{eq}) so that $\,c_\chi^2/(4\pi\,m_U^2)=c_{\chi\,p}^{\,2}/(16\pi\,m_\chi^2)\,$.
But such effects are not expected to be essential 
at freeze-out time, at least for moderate values of 
the coupling, as light dark matter velocities, 
$\,v\approx .4\,c$, are not small.

\vspace{1mm}
Should such an enhancement factor be present for the annihilation cross section, (\ref{sizecf}) 
would be turned into an inequality, 
$\,|c_\chi\,f_e|\,<\ ...\,$, \,and similarly for the expression of 
$\,B(\Upsilon\to\chi\,\chi)$ as written in (\ref{relbinv}).
The discussion further illustrates 
that there is no general way to predict
the invisible $\Upsilon\to\chi\chi$ branching fraction simply from the annihilation cross section of 
light dark matter at freeze-out time.

\vspace{-1.5mm}

\section{Conclusion.}

\vspace{-1mm}

Searching for invisible $\psi$ and $\Upsilon\,$ decays has long been identified and used 
as a way to search for new light particles, especially if they are 
more strongly coupled to standard model ones than through weak interactions.
Invisible branching fractions into neutrinos in the SM and neutralinos in the (N/n)MSSM, however,
are well below present experimental limits, 
even if light spin-0 particles are directly coupled to both quarks and neutralinos. 

\vspace{1mm}
Exchanges of a new heavy quark $\,b_M$ or $c_M$ also lead to small invisible branching fractions 
into scalar dark matter, unless the $\,b_M$ or $c_M$ quarks are only moderately heavy, 
and the corresponding Yukawa couplings to scalar dark matter rather large.

\vspace{1mm}

Exchanges of a light spin-1 $U$ boson associated with a new gauge interaction, with a vector coupling to quarks, 
may induce invisible decays of $\,\psi$ and $\,\Upsilon$ into light dark matter particles at a significant rate, or, conversely, at a very small rate,
with $\,U$ exchanges responsible for sufficient dark matter annihilations in the early Universe.

\vspace{.5mm}
The new limit on invisible $\Upsilon\,$ decays constrain the $U$ couplings 
to dark matter and $b$ quarks to satisfy $\,\,|\,c_\chi\,f_{bV}\,|\,<\,5\ 10^{-3}$, 
\,for $\,2\,m_\chi\,$  smaller than a few GeV's. This may be compared with 
$\,\,|\,c_\chi\,f_{cV}\,|\,<\,9.5\ 10^{-3}$ from BES II (expected to get significantly improved at BES III).
Small values of $m_\chi$ of a few MeV's, and  $\,m_U$ of less than a few hundred MeV's, 
in fact, may well be preferred.

\vspace{.5mm}

The radiative decays $\,\psi\,$ or $\Upsilon\to\gamma\,\chi\chi\,$, or $\,\gamma\,\varphi\,\bar\varphi\,$,
may be induced by the axial couplings of the $U$ to the $c$ and $b$ quarks, with amplitudes proportional to $\,c_\chi f_{cA}\,e$ or $\,c_\chi f_{bA}\,e$. 
If they can be sufficiently well constrained although the emitted photon is not monochromatic, 
they may lead to interesting limits on 
$\,c_\chi\,f_{cA}$ or $\,c_\chi\,f_{bA}$, \,especially for larger $\,c_\chi$,
as $f_{cA}$ and $f_{bA}$ are already constrained from $\,\psi\,$ or $\Upsilon\to\gamma\,U$.
These decays may also be sensitive to heavy-quark exchanges, for scalar dark matter.

\vspace{.5mm}

Even if standard decay modes into neutrinos are not reachable yet, 
searches for invisible meson decays give useful informations on the $U$ boson and its couplings, 
contributing to shed light on the nature of dark matter and on the dark force through which it 
interacts with ordinary particles.

\end{document}